\documentclass[11pt]{article}

\usepackage{lineno,hyperref}
\usepackage{graphicx}
\usepackage{natbib}

\title{A comparison of remotely-sensed and inventory datasets for burned area in Mediterranean Europe}

\author{Marco Turco$^1$, \and Sixto Herrera$^2$, \and Etienne Tourigny$^1$, Emilio Chuvieco$^3$ and \and Antonello Provenzale$^4$}
\date{%
    $^1$Earth Sciences Department, Barcelona Supercomputing Center (BSC), C/ Jordi Girona 29, 08034 Barcelona, Spain\\%
    $^2$Grupo de Meteorolog\'ia. Dpto. de Matem\'atica Aplicada y Ciencias de la Computaci\'on, Universidad de Cantabria, Avda. de los Castros, s/n. 39005. Santander. Spain \\
    $^3$Environmental Remote Sensing Research Group, Department of Geology, Geography and the Environment, Universidad de AlcalaÃÅ, Calle Colegios 2, AlcalaÃÅ de Henares, 28801, Spain \\
    $^4$Institute of Geosciences and Earth Resources, National Research Council, Via Moruzzi 101, 56124 Pisa, Italy \\
    \today
}

\begin{document}

\maketitle

\copyright 2019. Licensed under the Creative Commons 

[CC-BY-NC-ND https://creativecommons.org/licenses/by-nc-nd/4.0/].

\pagebreak
\newpage

\section{Abstract}

Quantitative estimate of observational uncertainty is an essential ingredient to correctly interpret changes in climatic and environmental variables such as wildfires. In this work we compare four state-of-the-art satellite fire products with the gridded, ground-based EFFIS dataset for Mediterranean Europe and analyse their statistical differences. The data are compared for spatial and temporal similarities at different aggregations to identify a spatial scale at which most of the observations provide equivalent results. The results of the analysis indicate that the datasets show high temporal correlation with each other (0.5/0.6) when aggregating the data at resolution of at least $1.0^\circ$ or at NUTS3 level. However, burned area estimates vary widely between datasets. Filtering out satellite fires located on urban and crop land cover classes greatly improves the agreement with EFFIS data. Finally, in spite of the differences found in the area estimates, the spatial pattern is similar for all the datasets, with spatial correlation increasing as the resolution decreases. Also, the general reasonable agreement between satellite products builds confidence in using these datasets and in particular the most-recent developed dataset, FireCCI51, shows the best agreement with EFFIS overall. As a result, the main conclusion of the study is that users should carefully consider the limitations of the satellite fire estimates currently available, as their uncertainties cannot be neglected in the overall uncertainty estimate/cascade that should accompany global or regional change studies and that removing fires on human-dominated land areas is key to analyze forest fires estimation from satellite products.


\section{keyword}
Burned area; observational uncertainty; climate analysis; Mediterranean basin; Earth Science Information

\section{Introduction}

Wildfires are often devastating in countries with dense urban-wildland interface: every year, thousands of people around the world are affected by fire, with a severe toll of human life losses and significant economic damage. For instance, during the summer of 2018, Australia, Greece, North America, Scandinavia (with some areas even within the Artic Circle) and the United Kingdom experienced unusually destructive wildfires. Close to Athens, a series of wildfires killed 99 people in July 2018, the deadliest fires in recent Greek history \citep{aghakouchak2018natural}. In addition, societal exposure to large fires has been increasing \citep{moritz2014learning, bowman2017human}. For these reasons, both the occurrence of and the changes in climatic extremes are of great concern for their impact on fire occurrence, as weather factors control fire severity and the extent of the burned area \citep{jolly2015climate, turco2018exacerbated}.

Reliable information on Burned Area (BA) is crucial to identify the drivers of changes in fire activity \citep{forkel2017data, andela2017human}, assess fire risk \citep{chuvieco2014integrating, lasaponara2018fire} and/or develop fire risk prediction tools \citep{turco2018skilful}. Also, observed BA data are used to validate dynamic global vegetation models \citep{lasslop2014spitfire}, as input for atmospheric emission models \citep{van2017global}, and for estimating fire impacts on human health \citep{reid2016critical} and safety of properties \citep{moritz2014learning}. It is also worth recalling that fire disturbance is one of the Essential Climate Variables (ECVs) defined by the Global Climate Observing System (GCOS) programme \citep{belward2016global} since there are strong links between climate and fire (see e.g. \citealp{williams2016recent, forkel2017data, abatzoglou2018global}). Developing systematic assessments of fire data and improving the use of satellite products in impact modelling has thus become a strategic topic in national and international climate programs (see e.g. the $Fire\_cci$ project; https://www.esa-fire-cci.org/, last access: May 2019). 

There is now a growing body of studies that use BA estimates based on satellite products \citep{mouillot2014ten, chuvieco2019historical}. The Moderate Resolution Imaging Spectroradiometer (MODIS), on board the Terra and Aqua satellites since 2000 (https://modis.gsfc.nasa.gov/about/, last access: May 2019), has been used to generate the global BA time series MCD64A1, now in its collection 6 \citep{giglio2018collection} and to derive other variables \citep[GFED;  ][that includes fuel properties and emission coefficients]{van2017global}. The $Fire\_cci$ project, which is part of the European Space Agency (ESA) Climate Change Initiative (CCI) \citep{hollmann2013esa}, has developed long-term time series of Burned Area products, the MODIS $Fire\_cci$ v5.1 Burned Area product (hereinafter FireCCI51), adapted to the needs of impact modellers \citep{chuvieco2016new, Chuvieco_et_al_2019}.

On the other hand, BA estimates vary widely between the different datasets \citep{mouillot2014ten, chuvieco2019historical} and, thus, rigorous evaluation of BA estimates is a much needed step to assess the reliability of the information. Unfortunately, a number of constraints may limit such analysis (see \citealp{mouillot2005fire} for a global synthesis on national fire statistics and their limitations). In the case of ground-based BA data, they are commonly affected by:

\begin{itemize}
\item periods of unavailable or poor-quality data \citep{mouillot2005fire, koutsias2013relationships};
\item difficulties in the estimation of Burned Area from field observations \citep{pereira2011history, short2015sources};
\item uncertainties due to different fire report protocols between countries and/or protocol changes during time \citep{turco2013decreasing};
\item high political controls on BA statistics \citep{kasischke2000direct, belhadj2018revised};
\end{itemize}

On the other hand, satellite products provide global information but they cover only the last two decades. In addition, most assessments of satellite-derived fire datasets are based on comparisons between different satellite products (see e.g. \citealp{Chuvieco_et_al_2018, humber2018spatial}), with scarce attention to comparing satellite BA estimates with independent ground-based observations. 

Previous studies have shown that MODIS products are suitable proxies for national statistics in North America, Russia and China \citep{chang2009comparison, giglio2010assessing}, but large discrepancies emerge when comparing them to higher resolution \citep{fusco2019detection}. At the moment, it remains unclear what is the temporal-spatial aggregation scale at which satellite BA products are more consistent and reliable. In Europe, two studies compare remote sensing and field BA data at national scale \citep{loepfe2012comparison, vilar2015comparison}, finding high correlation between the MODIS BA product and the European national statistics, with a slight underestimation of the total BA. However, the national scale of these studies prevents local interpretation. 

To summarise, a comprehensive evaluation of satellite BA products with official statistics in Mediterranean Europe at high spatial resolution remains undone. Indeed, Mediterranean Europe, where the analysis of \citet{vilar2015comparison} is focused, is a crucial region for a detailed comparison study, owing to both the large impact of fires across the area \citep{San-Miguel-Ayanz2013} and the availability of a highly-controlled fire dataset based on ground observations. The database of the European Forest Fire Information System (EFFIS; \citealp{San-Miguel-Ayanz_et_al_2012}) is available at monthly temporal resolution and high spatial resolution for a long period (1985-2011) for Portugal, Spain, Southern France, Italy and France. This high-quality fire dataset, even though it cannot clearly be taken as error-free ``ground truth'', represents the state-of-the-art information available today at the scale of Mediterranean Europe and it calls for further analysis, allowing larger confidence in the results of the comparison with satellite products. 

Responding to such challenge, this paper presents a spatio--temporal comparison between the ground-based BA dataset from EFFIS and BA estimations obtained from several state-of-the-art satellite products (MODIS, GFED4, GFED4s and FireCCI51). This paper is structured as follows: in Section \ref{s.methods} a description of the data and methods considered in this work is presented. The main results are described in Section \ref{s.results}. The main conclusions derived from the analysis are reported in Section \ref{s.conclusions}.
 
 
\section{Methods}\label{s.methods}

\subsection{Data}

\begin{figure}[ht]
\begin{center}$
\begin{array}{c}
\noindent\includegraphics[width=0.9\textwidth]{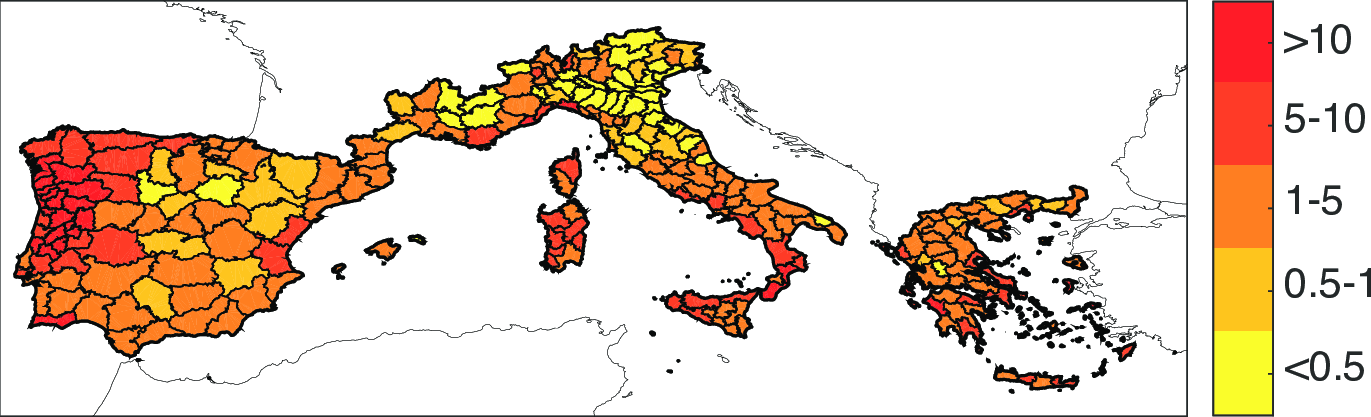}  \\
\end{array}$
\end{center}
\caption{Annual mean burned area fraction in southern Europe for the 1985--2011 period (in permillage). }\label{f.fig01}
\end{figure}

The high-quality database provided by the European Forest Fire Information System \citep[EFFIS]{San-Miguel-Ayanz_et_al_2012} has been compiled by the Joint Research Centre and the Directorate--General for the Environment of the European Commission, and it is the main source of harmonized data on forest fires in Europe. From EFFIS we obtained monthly BA data at the NUTS3 level (Nomenclature of Units for Territorial Statistics, which corresponds to aggregations of municipalities or provinces) for Portugal (1980-2015), Spain (1985-2014), southern France (1985-2016) and Italy (1985-2015), and Greece (1983-2011). The NUTS classification has been modified several times since its implementation in 2003. The EFFIS data are provided for the NUTS3 2006 version (see \url{http://ec.europa.eu/eurostat/web/nuts/} for more details, last access: May 2019), with a typical resolution of 10000-15000 km$^2$, corresponding to a grid-box with size of about 1-1.2 degrees. The EFFIS dataset comes from the national member states fire statistics and it consists of BA occurring in forests and other land areas, excluding agricultural or other artificial surfaces \citep[as detailed in ][]{San-Miguel-Ayanz_et_al_2012}. Figure \ref{f.fig01} shows the long-term average over the common period (1985-2011) of the EFFIS data at NUTS3 level and indicates that the spatial variability of annual BA is quite high, with highest values in the northwestern and eastern Iberian Peninsula, as well as in Southern Italy and in Greece. EFFIS data have been widely used in studies that analyse spatial-temporal fire changes and identify fire drivers (see e.g. \citealp{turco2017key}). 

For the comparison, we focused on four satellite products providing monthly burned area information:

\begin{itemize}
\item the Moderate Resolution Imaging Spectroradiometer MCD64A1 c6 (hereinafter MODIS; \citealp{giglio2018collection});
\item the fourth generation of the Global Fire Emissions Database (GFED4; \citealp{Giglio_et_al_2013});
\item the GFED4s, which extends the GFED4 including the BA from ``small" fires. Specifically, GFED4s includes an estimation of the area burned by small fires ($<$ 100 ha) based on MODIS active fire hotspots located outside of burned patches mapped in the MCD64A1 BA product. The BA allocated to each outside-of-burn hotspot is in turn estimated using postulated relationships between the difference normalized burn ratio, the number of within-burn hotspots, and the BA actually mapped in the MCD64A1 product. See \cite{Randerson_et_al_2012} for more details;
\item the MODIS $Fire\_cci$ v5.1 Burned Area product (FireCCI51), based on the highest spatial-resolution bands of the MODIS sensor (R, red, and NIR, near-infrared) and complementing existing global BA datasets using higher spatial-resolution bands \citep{Chuvieco_et_al_2018, Chuvieco_et_al_2019}.
\end{itemize}

Table~\ref{t.datasets} summarizes the Burned Area data used in the study, their spatial and temporal availability, resolution and providers.

\begin{table}[htb]
\caption{Description of the datasets used in this study. *EUMED domain corresponds to Portugal, Spain, Southern France, Italy and Greece.}\label{t.datasets}
\begin{center}
\begin{tabular}{l l l l}
\emph{Dataset} & \emph{Coverage(T/S)} & \emph{Resolution(T/S)} & \emph{Source}\\\hline
EFFIS & 1985-2011/EUMED* & month/NUTS3 & \url{http://effis.jrc.ec.europa.eu/}\\
FireCCI51 & 2001-2017/Global & month/$0.25^{\circ}$   & \url{https://www.esa-fire-cci.org/} \\
GFED4 & 1995-2016/Global & month/$0.25^{\circ}$ & \url{http://www.globalfiredata.org} \\
GFED4s & 1997-2016/Global & month/$0.25^{\circ}$ &  \url{http://www.globalfiredata.org}  \\
MODIS & 2001-2016/Global & month/$0.25^{\circ}$ & \url{http://modis-fire.umd.edu/} \\
\hline
\end{tabular}
\end{center}
\end{table}



Although some countries have data also after 2011, unfortunately for Greece there are only data until this year. Thus, to obtain a consistent and homogeneous ensemble, only the common period, 2001-2011, has been considered in the comparison. The base spatial resolution considered in the analysis is $0.25^{\circ}$ at monthly temporal resolution. The MODIS products were downloaded at their nominal spatial resolution (500m), and were then aggregated at the $0.25^{\circ}$ resolution for comparison with the other products.

Since EFFIS data are focused on report forest fires, i.e. excluding agricultural or pasture fires, we performed a filtering of the satellite data (when possible) according to the land cover. Then we analyze the sensitivity of the results of this filtering by comparing unfiltered and filtered data. Specifically, we filter out from the FireCCI51 BA data the values from the land categories 'cropland, rainfed', 'cropland, irrigated or post-flooding' and 'mosaic cropland ($>$50\%)'. In the following we refer to this data as FireCCI51-nat. GFED4 data provide the land cover distribution of burned area in percentage. We filter out the GFED4 data in case the corresponding land cover distributions 'croplands' or 'urban and built-up' are greater then 50\%. In the following we refer to this data as GFED4-nat. Similarly, we filtered the MODIS data considering the land cover 'croplands' and 'urban and built-up' according to the MODIS Land Cover Type Product (MCD12Q1). Hereinafter we refer to this data as MODIS-nat.

\subsection{Comparison methodology}


To have a complete view of the resolution-dependent differences between datasets at different spatial scales, we follow two approaches. First, in addition to the base resolution of $0.25^{\circ}$, all datasets have been conservatively aggregated also at $0.5^{\circ}$, $1.0^{\circ}$, $1.5^{\circ}$, $2.0^{\circ}$ and $2.5^{\circ}$ resolutions. To perform such analysis, the NUTS3-level EFFIS dataset was downscaled at the resolution of the satellite products ($0.25^{\circ}$), applying the following conservative procedure: 
\begin{itemize}
\item for each NUTS3 region we find the number, $n$, of pixels with resolution $0.25^{\circ}$ in this region, and 
\item we assign to such pixels the value of the BA given by EFFIS, corresponding to the burned area in that NUTS3 region, divided by $n$. In this way we ensure that the total BA of the NUTS3-level data is conserved.
\end{itemize}
This procedure imposes homogeneity of the BA estimates from EFFIS at scales smaller than those of the NUTS3 regions, while satellite products display natural variability on these scales. Thus, caution should accompany the comparison between satellite products and EFFIS at these smaller scales. On the other hand, satellite data can be compared with each other also at scales smaller than NUTS3.

As an alternative approach, we also compared the BA values within the NUTS classification. This is a hierarchical way for aggregating the territory of Europe: Specifically, we compared the datasets aggregating their values over NUTS3 (small regions), NUTS2 (basic regions that aggregated NUTS3 area) and NUTS1 areas (major socio-economic regions that aggregated NUTS2 areas). Finally, following the approach of previous studies, the accumulated value of BA over the entire domain has also been considered in the analysis.

The temporal and spatial statistical differences/similarities between the satellite products (called SAT) and EFFIS, considering the latter as ground reference, have been evaluated for each resolution. For satellite products, for each aggregation scale, we estimated the mean relative error ($ME_{r} = 100*\frac{\overline{SAT-EFFIS}}{\overline{EFFIS}}$) and the Pearson correlation of the monthly series (Spearman correlation provides very similar results, not shown).

To compare the spatial pattern, we assessed the agreement of the mean values (averaged over the common period 2001-2011) of the annual BA for the EFFIS data and satellite products. The spatial pattern obtained for each satellite products is compared to EFFIS data using the correlation, the variability (standard deviation), the centred root-mean-square error and the bias. To make the different indices comparable, standard deviation and centred root-mean-square error statistics have been normalized by dividing for the standard deviation of the reference EFFIS dataset, and the bias has been normalized with respect to the mean of the reference dataset. Finally, Taylor diagrams \citep{taylor2001summarizing} have been used to summarize these metrics on a single plot.

\clearpage
\newpage	

\section{Results of the analysis}\label{s.results}

\begin{figure}[ht!]
\begin{center}$
\begin{array}{c}
\noindent\includegraphics[width=0.9\textwidth]{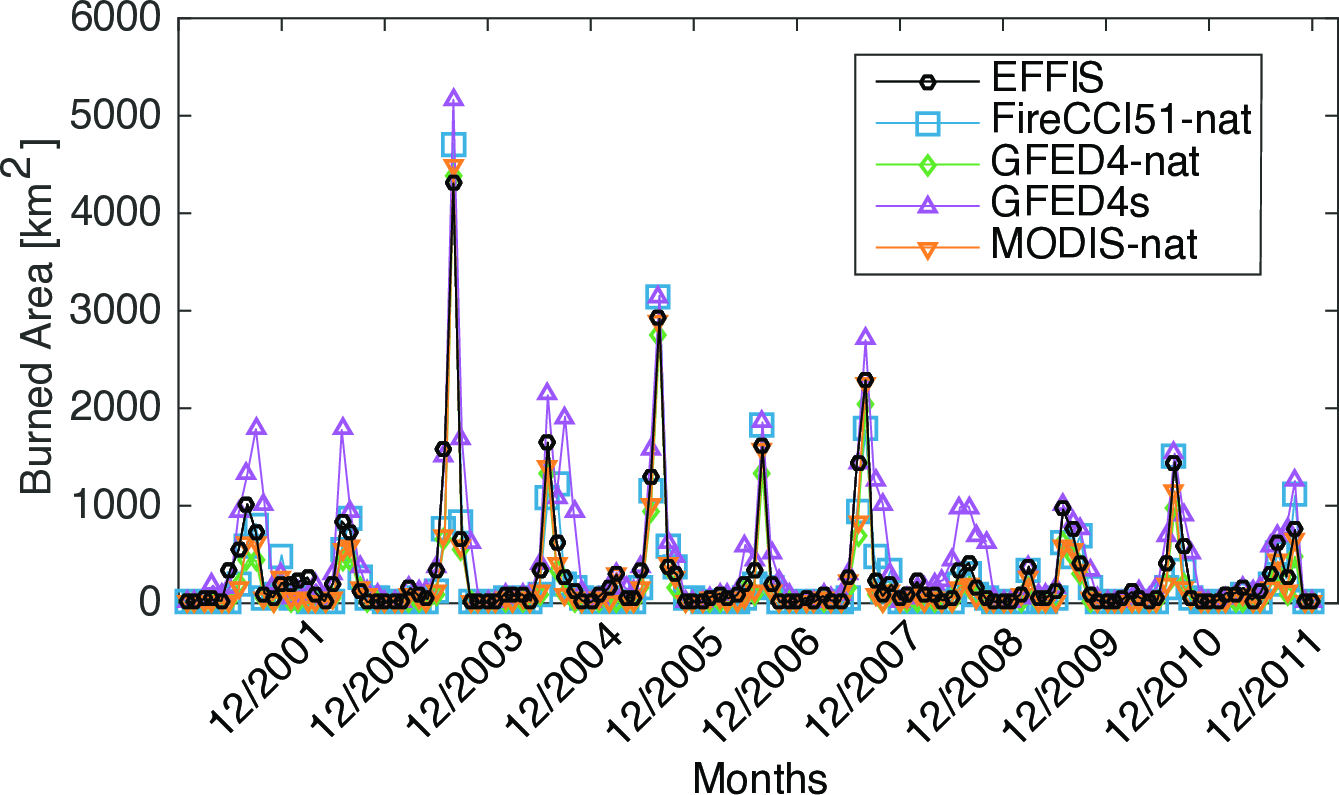}  \\
\end{array}$
\end{center}
\caption{Monthly burned area in southern Europe (in km$^2$) for each dataset.}\label{f.fig02}
\end{figure}

\subsection{Temporal correlations}

We first estimate the temporal consistency of the five datasets. Figure \ref{f.fig02} shows the total monthly BA over the entire domain (see Figure \ref{f.fig01}), that is, summing the monthly BA values over the whole Mediterranean Europe. A strong periodicity, with the highest values during summer months, is evident in all datasets. Table~\ref{t.correlations} indicates strong agreement of the datasets at this large spatial scale, with correlations ranging from $0.92$ to $0.97$ (third column in Table~\ref{t.correlations}). To distinguish the interannual variability from the annual cycle, the correlations of monthly anomalies (obtained by subtracting the mean annual cycle; fourth column in Table~\ref{t.correlations}) were also estimated, obtaining values ranging from $0.92$ to $0.97$. This confirms the excellent temporal correlation between datasets at the scale of Mediterranean Europe. 

\begin{table}[htb]
\caption{Temporal consistency between the datasets for the whole EUMED domain. The first column indicates the dataset. The second column shows the annual mean total burned area. The third and fourth columns report respectively the correlation between the satellite datasets and the EFFIS monthly data, and between the monthly anomalies, obtained by subtracting the mean annual cycle.}\label{t.correlations}
\begin{center}
\begin{tabular}{l c c c}
\emph{Dataset} & \emph{Burned area ($km^2$)} & \emph{Monthly correlations} & \emph{Monthly anomalies correlations}\\\hline
EFFIS & 3597 & 1 & 1\\
FireCCI51-nat & 3085 (2289-3729) & 0.96 & 0.94 \\
FireCCI51 & 6187 (5467-6907) & 0.94 & 0.90 \\
GFED4-nat & 2204 (867-3597) & 0.97 & 0.97 \\
GFED4 & 3291 (1020-5564) & 0.97 & 0.95 \\
GFED4s & 5632 & 0.92 & 0.90 \\
MODIS-nat & 2.557 & 0.97 & 0.97 \\
MODIS & 4515 & 0.95 & 0.92 \\
\hline
\end{tabular}
\end{center}
\end{table}

Although the temporal correlation between the BA series is very high, the total BA values, averaged over the common period 2001-2011, display relevant differences between the different datasets (Table~\ref{t.correlations}). EFFIS data indicate that on average, about 3600 $km^2$ are burned every year. The closest estimation to this value is provided by the FireCCI51-nat dataset, with about 3100 $km^2$ (i.e. with a difference of about $-14\%$). The standard error field provide by this dataset indicated that, although the spread is quite large (2289-3729), it includes the EFFIS estimate. Also, it is worth noting that for the unfiltered FireCCI51 data, the BA values over the region is much higher, according to previous studies indicating that over Europe many fires occur over cropland areas \citep{Giglio_et_al_2013}. The GFED4-nat dataset shows a lower value ($-39\%$), with a larger spread 867-3597 and shows more similar estimate to the EFFIS one considering the unfiltered data. Interestingly, the 'nat' datasets improves also the correlation values. The MODIS unfiltered data show larger values ($+25\%$) and GFED4s even larger differences ($56\%$). Unfortunately, these two datasets do not provide information on the BA uncertainty and the GFED4s data does not provide information on the land cover data. Small fires often occur in pastures, agricultural lands and in other landscapes dominated by the human presence \citep{Randerson_et_al_2012, van2017global}. Thus the large difference between GFED4s and EFFIS might be explained since the former includes BA from ``small" fires while the latter includes fires usually referenced for forest or wildland areas, i.e. excluding fires in agricultural setting.



\begin{figure}[ht]
\begin{center}$
\begin{array}{c}
\noindent\includegraphics[width=0.9\textwidth]{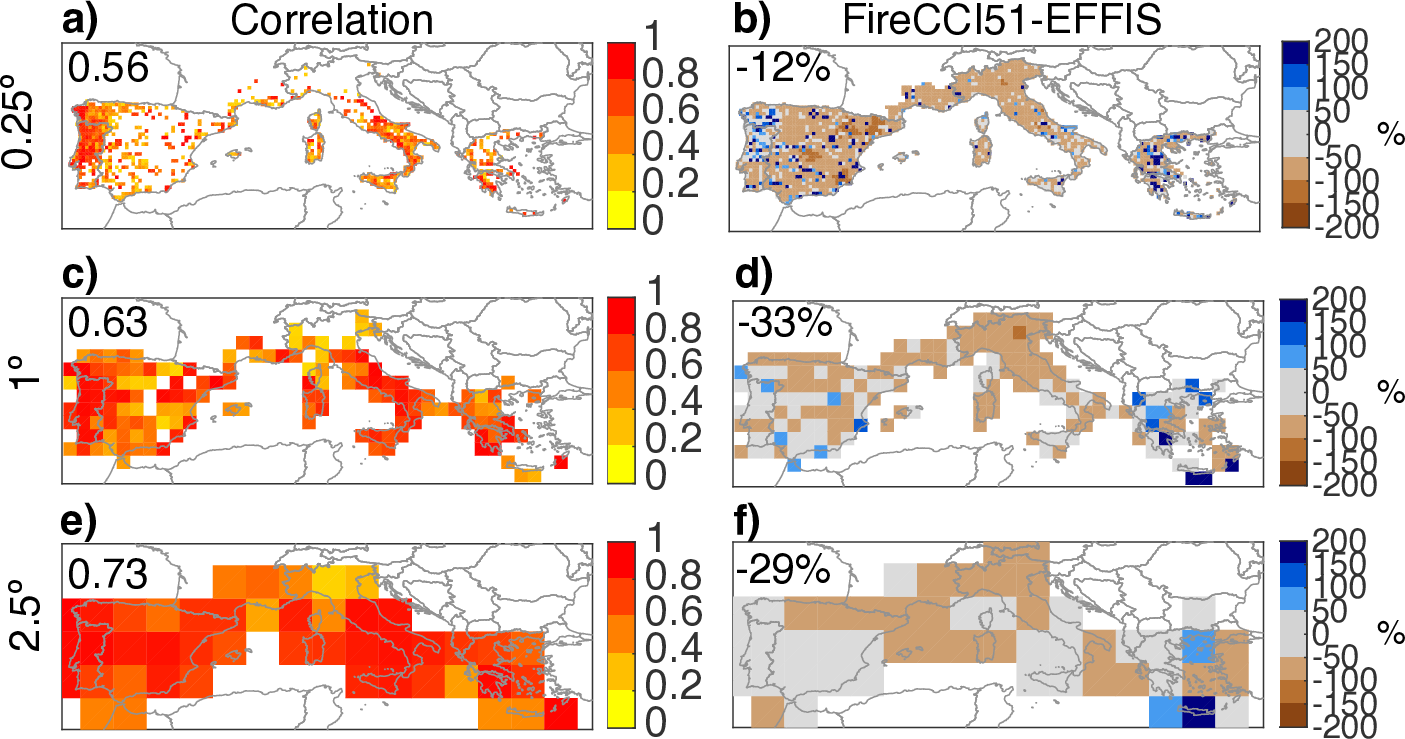}  \\
\end{array}$
\end{center}
\caption{Pearson correlation (left) and $ME_{r}$ (right) comparing EFFIS and FireCCI51-nat at $0.25^\circ$ (a-b),  $1.0^\circ$ (c-d) and  $2.5^\circ$ (e-f) resolutions. Only significant ($p-value < 0.05$) correlations are shown. The value inside each panel corresponds to the spatial mean of the score (in case of correlation, only statistically significant values have been considered in the average).}\label{f.fig03}
\end{figure}

\begin{figure}[ht]
\begin{center}$
\begin{array}{c}
\noindent\includegraphics[width=0.9\textwidth]{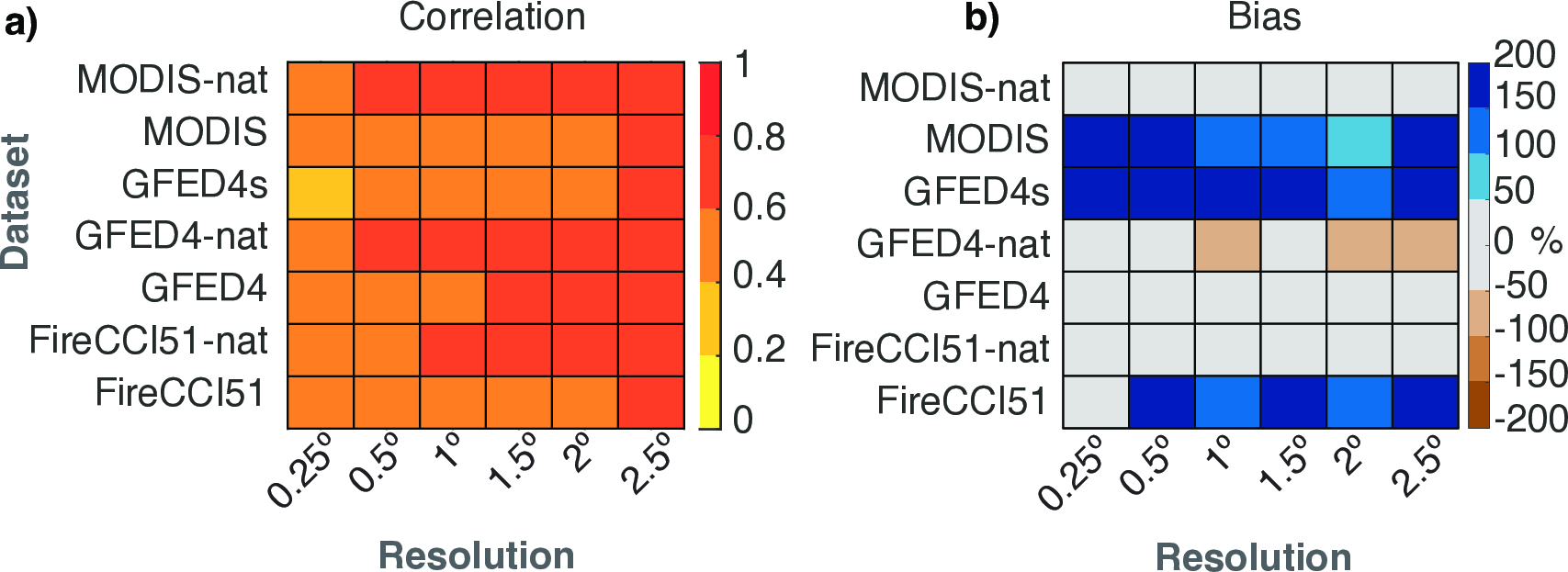}  \\
\end{array}$
\end{center}
\caption{Spatial mean of (a) Pearson correlation (considering only pixels with statistically significant correlations) and (b) $ME_{r}$ for different satellite products and resolutions.}\label{f.fig04}
\end{figure}

To identify the spatial scales where the datasets display the stronger agreement, we first focus on different spatial aggregations of the data and compare decreasing resolutions from $0.25^\circ$ to $2.5^\circ$. As an illustrative example, Figure \ref{f.fig03} shows the correlation and the difference between the FireCCI51 and the EFFIS data. Correlation values increase as the resolution decreases, with spatially averaged values equal to $0.56$ at $0.25^\circ$ and to $0.73$ at $2.5^\circ$. At this coarser resolution, most of the domain pixels display significant correlations between the two datasets. On average the bias of FireCCI51 with respect to EFFIS is negative, with few areas with positive differences, mainly over Greece. Overall, at $2.5^\circ$ resolution, most of the grids show differences lower than 50\%. Figure \ref{f.fig04} summarizes the temporal similarity scores for the different resolutions (0.25, 0.5, 1, 1.5, 2 and 2.5$^\circ$) and the satellite products (filtered or not). Figure \ref{f.fig04}(a) confirms that the correlation values are larger considering the filtered products ("nat") and for coarser resolutions. On the other hand, Figure \ref{f.fig04}(b) indicates that the filtered products generally have a lower bias with respect the original datasets, and that these ones show higher values than the EFFIS data. Overall, MODIS-nat and FireCCI51 show the best agreement with the EFFIS data.

\begin{figure}[ht]
\begin{center}$
\begin{array}{c}
\noindent\includegraphics[width=1.1\textwidth]{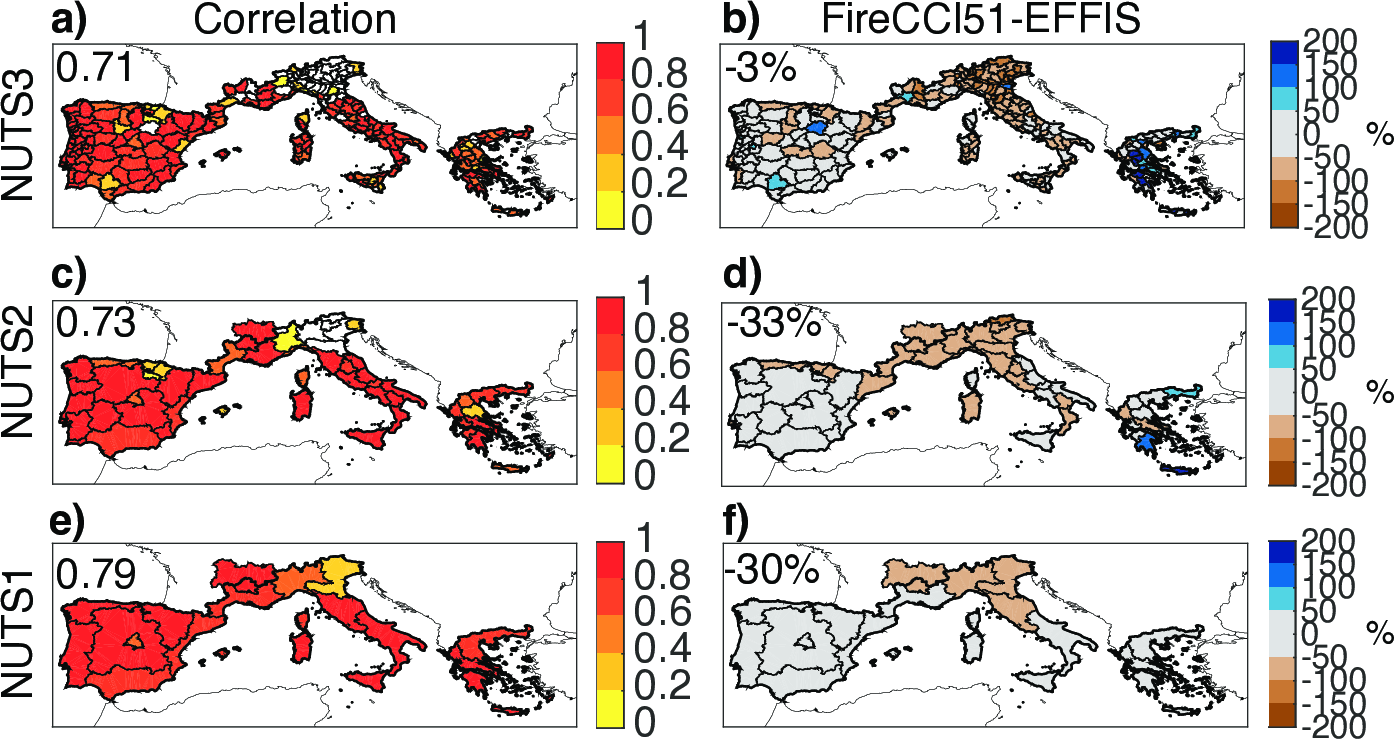}  \\
\end{array}$
\end{center}
\caption{Pearson correlation (left) and $ME_{r}$ (right) comparing EFFIS and FireCCI51-nat at NUTS3 (a-b),  NUTS2 (c-d) and  NUTS1 (e-f) level. Only significant ($p-value < 0.05$) correlations are shown. The value inside each panel corresponds to the spatial mean of the score (in case of correlation, only statistically significant values have been considered in the average).}\label{f.fig05}
\end{figure}

Following the second approach mentioned above, Figure \ref{f.fig05} shows the comparison between the FireCCI51-nat and EFFIS data within NUTS3, NUTS2 and NUTS1 regions and indicates that at the larger scale of aggregation (NUTS1), all the domain show statistically significant correlation, with spatially averaged value of 0.79 and a difference of -30\%. Interesting, the results are similar also aggregating the data at NUTS2 and NUTS3 level. 

\begin{figure}[ht]
\begin{center}$
\begin{array}{c}
\noindent\includegraphics[width=0.9\textwidth]{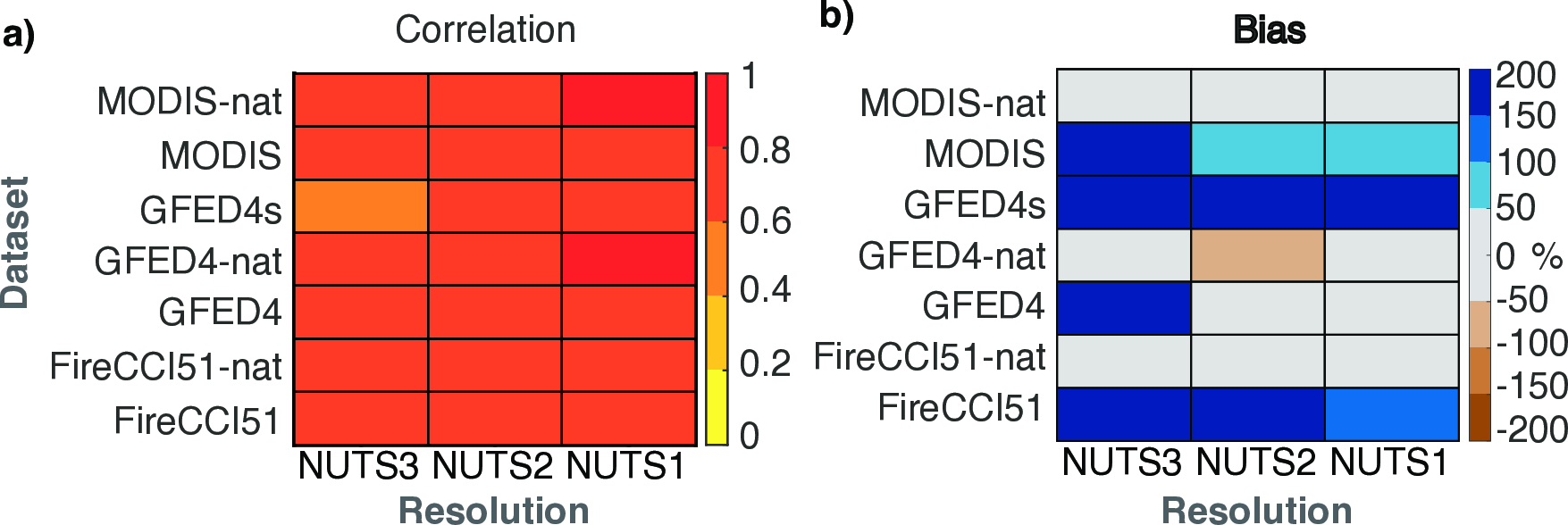}  \\
\end{array}$
\end{center}
\caption{Spatial mean of (a) Pearson correlation (considering only pixels with statistically significant correlations) and (b) $ME_{r}$ for different satellite products and NUTS aggregations.}\label{f.fig06}
\end{figure}

Considering all the datasets and the NUTS divisions (Figure \ref{f.fig06}), three main conclusion can be drawn. First, the best correlation values are obtained considering the NUTS1 regions, with spatially averaged values larger then 0.7. Second, generally larger correlations are obtained considering the "nat" datasets. Finally, in term of difference ("bias" in Figure \ref{f.fig06}), this analysis confirms that higher similarity between EFFIS and remotely-sensed datasets are obtained considering the "nat" versions of the latter.

\subsection{Spatial similarity}

\begin{figure}[ht]
\begin{center}$
\begin{array}{c}
\noindent\includegraphics[width=1.1\textwidth]{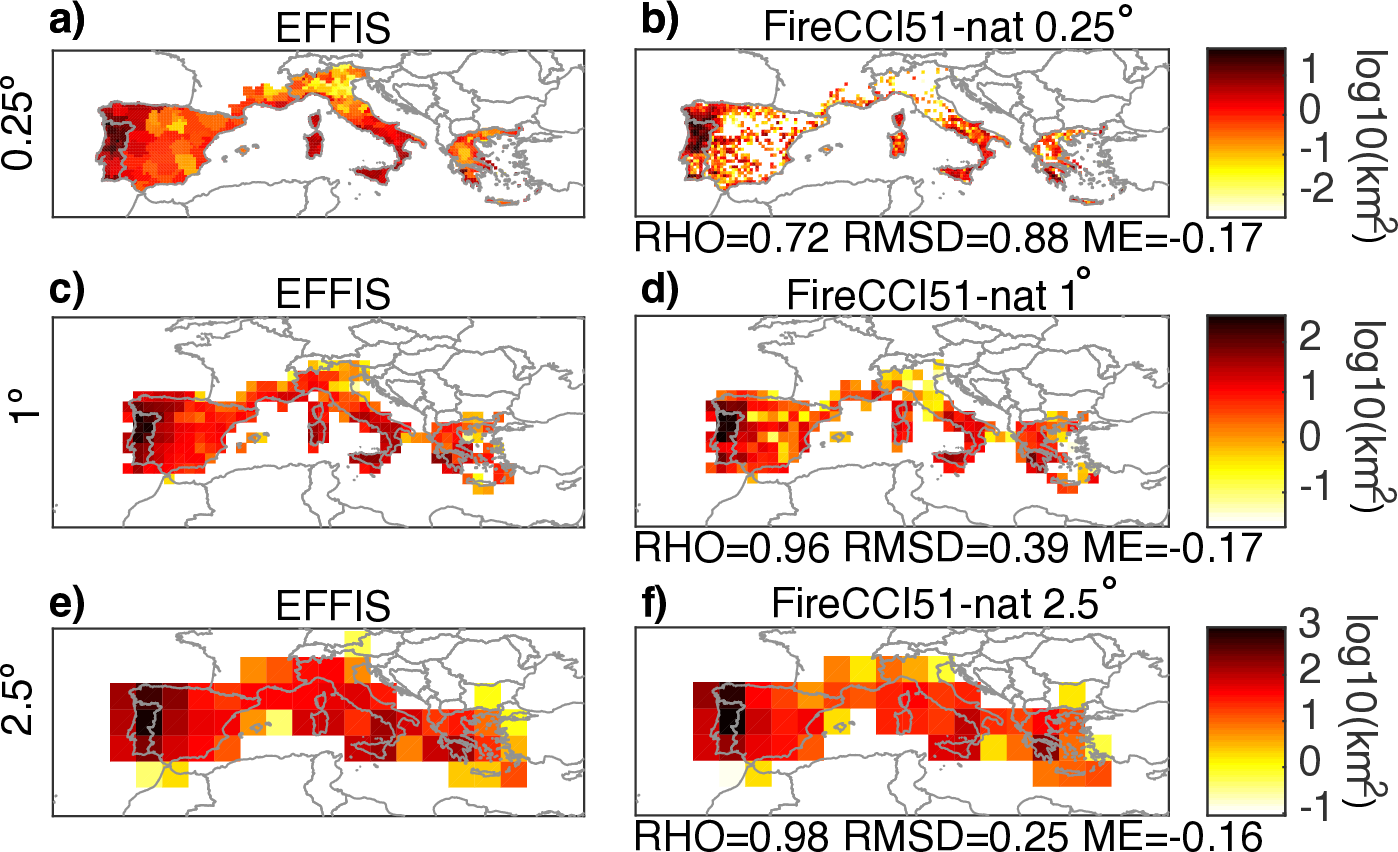}  \\
\end{array}$
\end{center}
\caption{a) Spatial distribution of annual BA series, averaged over the period 2001-2011, for a) EFFIS and b) FireCCI51-nat data at resolution 0.25$^\circ$. Panels c) and d) are the same as a) and b), respectively, for resolution 1$^\circ$. Panels e) and f) are the same as a) and b), respectively, for resolution 2.5$^\circ$. The spatial validation scores (correlation RHO, centred root mean square RMSD, and mean error ME) for the FireCCI51-nat values, with respect to the EFFIS values, are given below the corresponding panels.}\label{f.fig07}
\end{figure}

Figure \ref{f.fig07} shows the mean values (averaged over the common period 2001-2011) of the annual BA for the EFFIS and FireCCI51 datasets. This comparison reveals the similarity between these datasets, with spatial correlation above 0.9, bias $<$ -$20\%$, and a centred root-mean-square error of less than 0.4 for aggregation larger then $1.0^{\circ}$. At finer resolution ($0.25^{\circ}$), the agreement is lower but still quite good, with spatial correlation around 0.7. As mentioned above, the lower agreement at scales lower than the NUTS3 (about 1 square degree) area can also be due to the homogeneous interpolation of the EFFIS data. Higher resolution ground data (currently unavailable from EFFIS over the Mediterranean Europe domain) would be needed to validate satellite fire products at smaller scale. 

\begin{figure}[ht]
\begin{center}$
\begin{array}{c}
\noindent\includegraphics[width=0.6\textwidth]{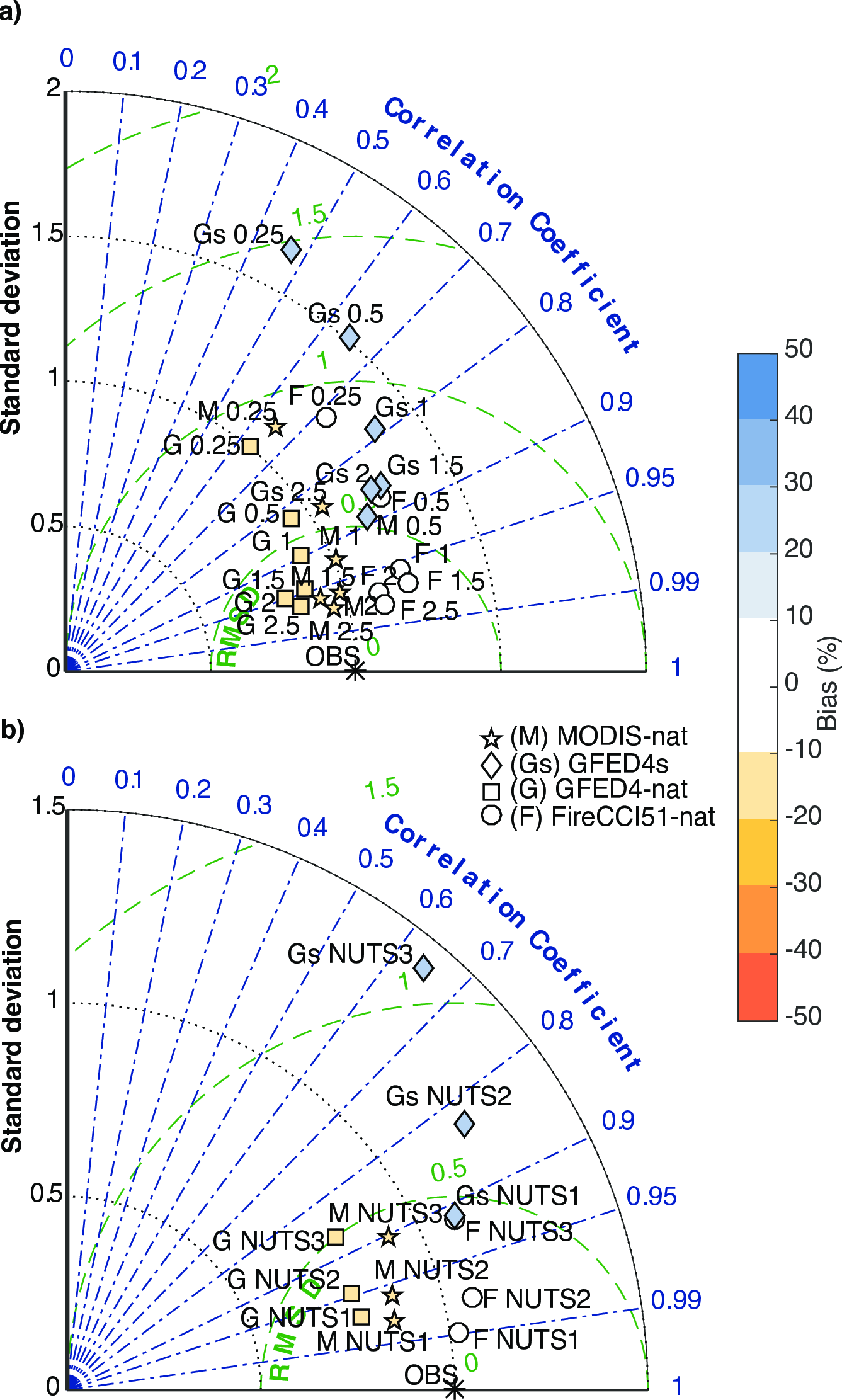}  \\
\end{array}$
\end{center}
\caption{Taylor diagrams summarising the spatial similarity to EFFIS of the mean annual burned series for different satellite products and a) resolutions and b) NUTS aggregations. The products that are more similar to the reference (EFFIS) are closer to the asterisk indicated as OBS. The colours indicate the bias (in percentage with respect to the EFFIS data) of the dataset. The numbers correspond to the different resolutions while the letters and the markers correspond to the different datasets: circle--F is FireCCI51-nat, square--G is GFED4-nat, diamond--Gs is GFED4s and star--M is MODIS-nat.}\label{f.fig08}
\end{figure}

The Taylor diagrams in Figure \ref{f.fig08} show a summary of the scores for all satellite products and type of aggregations (grids or NUTS). In the figure, the satellite datasets that are more similar to the EFFIS data are close, in Taylor diagram space, to the observations (labelled as OBS). The diagrams confirm that generally higher spatial agreement is obtained with lower resolution and the closest agreement with EFFIS is obtained for the MODIS-nat, GFED4-nat and FireCCI51-nat products. For instance, these datasets, at resolution $2.5^{\circ}$ or at NUTS1 level, display a spatial correlation close or above 0.95, a bias lower then 10\%, a relative RMSD lower then 0.5 and a standard deviation of around 1, that is, a  spatial variability very close to that of the EFFIS data.


\clearpage
\newpage	

\section{Conclusions}\label{s.conclusions}

In this work we provided an assessment of satellite-based Burned Area datasets in Mediterranean Europe (FireCCI51, GFED4, GFED4s and MODIS), comparing them with the ground-based EFFIS dataset. The results indicate that the satellite products generally show good agreement with EFFIS in terms of temporal correlation (correlation values above 0.9), when considering the total BA in the whole Mediterranean region. However, even at this sub-continental scale, large biases have been found for MODIS and GFED4s, with an overestimation compared to EFFIS of +25\% and +56\% respectively. Higher agreement has been found removing agricultural fires. The temporal correlation between the datasets decreases at finer aggregation scales, with reasonably good scores at resolution coarser than $1^\circ$. At higher resolution, also our interpolation of the EFFIS dataset (originally provided at NUTS3 level) can generate discrepancies with the satellite-based estimates. Unfortunately, no comprehensive ground-based fire dataset at higher resolution is currently available for the whole Mediterranean Europe.  

In the comparison of the spatial BA pattern between satellite products and the EFFIS BA series, temporally averaged over the common period 2001-2011, the results of the analysis indicate that the highest discrepancies were found considering spatial pattern with resolution $0.25^\circ$. The spatial patterns become more similar at lower resolution, and the satellite datasets generally show good agreement with EFFIS data considering spatial aggregation of a least $1^\circ$ or aggregating the data at NUTS3 level, a fact that provides confidence in using these products at this scale. The fact that the BA estimations are more similar when aggregating the data over larger areas might be explained by the fact that at coarser resolution some of the temporal and spatial noise apparent in the higher resolution data are reduced. At smaller scale, some caution should be adopted in drawing conclusions based only on the satellite datasets. In general, better agreement with EFFIS has been obtained for the MODIS, GFED4 and for FireCCI51 products. In particular the most-recent developed dataset, FireCCI51, show the best agreement with EFFIS overall. 

In practical applications, users should carefully take into account the limitations of the satellite products (as well as of any other dataset including EFFIS). Without proper high-resolution ground-based BA information it is difficult or impossible to properly document and validate fire patterns, as well as to analyse the causes and impacts of fires. The results of this work also emphasize the crucial importance of making environmental data accessible for research activities and application. Based on the results reported here, future work will consider the long-term predictability of climate-driven impacts on fires by means of a hybrid modelling strategy (see e.g. \citep{ceglar2018land}. This study is the first step in this direction, providing the groundwork to choose the most reliable datasets and products to develop seasonal-to-multiannual fire predictions for Mediterranean Europe.

\section{Acknowledgements}

The authors thank EFFIS (European Forest Fire Information System of the European Commission Joint Research Centre, http://effis.jrc.ec.europa.eu) for providing access to fire series EFFIS. M.T. and E.T. have received funding from the European Union's Horizon 2020 research and innovation programme under the Marie Sk\l{}odowska-Curie grant agreement No. 740073 (CLIM4CROP project) and grant agreement No. 748750 (SPFireSD project), respectively. The work of A.P. has been supported by the European Union's Horizon 2020 ECOPOTENTIAL project (grant agreement No. 641762).

\bibliographystyle{apalike} 

\end{document}